# Ion irradiation effects on a magnetic Si/Ni/Si trilayer and lateral magnetic-nonmagnetic multistrip patterning by focused ion beam

B. N. Dev[1,*], N. Banu[1], J. Fassbender[2], J. Grenzer[2], N. Schell[2,$], L. Bischoff[2], R. Groetzschel[2,#], J. McCord[3]

[1]Department of Materials Science, Indian Association for the Cultivation of Science, Jadavpur, Kolkata, India
[2]Institute of Ion-Beam Physics and Materials Research, Helmholtz Centre Dresden-Rossendorf, P.O. Box 510119, 01314 Dresden, Germany
[3]Institute of Materials Science, Kiel University, Kaiserstr. 2, 24143 Kiel, Germany

Abstract

Fabrication of a multistrip magnetic/nonmagnetic structure in a thin sandwiched Ni layer [Si(5 nm)/Ni(10 nm)/Si] by a focused ion beam (FIB) irradiation has been attempted. A control experiment was initially performed by irradiation with a standard 30 keV Ga ion beam at various fluences. Analyses were carried out by Rutherford backscattering spectrometry, X-ray reflectivity, magnetooptical Kerr effect (MOKE) measurements and MOKE microscopy. With increasing ion fluence, the coercivity as well as Kerr rotation decreases. A threshold ion fluence has been identified, where ferromagnetism of the Ni layer is lost at room temperature and due to Si incorporation into the Ni layer, a $Ni_{0.68}Si_{0.32}$ alloy layer is formed. This fluence was used in FIB irradiation of parallel 50 nm wide stripes, leaving 1 μm wide unirradiated stripes in between. MOKE microscopy on this FIB-patterned sample has revealed interacting magnetic domains across several stripes. Considering shape anisotropy effects, which would favor an alignment of magnetization parallel to the stripe axis, the opposite behavior is observed. Magneto-elastic effects introducing a stress-induced anisotropy component oriented perpendicular to the stripe axis are the most plausible explanation for the observed behavior.

1. Introduction

Magnetic patterning by means of ion irradiation and implantation has been an important area of research for the past couple of decades [1, 2, 3]. Such patterning has the potential for applications in magnetic data storage and magneto-logic devices. In ultrathin magnetic films and multilayers, owing to their reduced dimensions, the magnetic properties, e.g. magnetic anisotropies and exchange coupling crucially depend on surface and interface structure. Ion irradiation can modify these properties. For example, ion irradiation induced magnetization reorientation occurs in magnetic multilayers, e.g. Co/Pt multilayers. Energetic ions induce interface mixing, thereby triggering a spin reorientation transition [4]. Such systems are suited for patterned ultrahigh-density recording media. For similar intermixing at the interfaces, ion irradiation induced modification of exchange bias is observed in systems involving a ferromagnetic/antiferromagnetic interface [1] or even in a system containing nonmagnetic/magnetic interfaces, e.g. a Si/Co/Si trilayer [5]. Ion irradiation induced exchange bias modification is widely investigated for applications in spin-valves and magnetic tunnel junctions [1]. These modifications are predominantly due to ion beam induced intermixing at the interfaces. However, there are interesting alterations of magnetic behavior also due to ion beam induced atomic migration and diffusion [6,7].

Oscillatory magnetic coupling and giant magnetoresistance (GMR) effect were discovered in magnetic multilayers [8,9]. Results from basic research have rapidly gone into applications. Magnetic multilayer structures form an important component in today's computers as magnetic read head. GMR is displayed by a wide variety of nanostructures consisting of magnetic layers, separated by thin nonferromagnetic metallic spacer layers. A classic example of a magnetic multilayer is a Co/Cu periodic multilayer, which shows oscillatory magnetic coupling [10]. Here Co/Cu/Co/Cu layers are stacked along the depth of the multilayer and the nature of magnetic coupling, ferromagnetic or antiferromagnetic, depends on the thickness of the Cu spacer layers. As a logical extension of this system one can ask the question: Instead of a multilayer system, if one deals with a single layer, which is laterally patterned in a way so that alternating strips of materials in the layer are Co and Cu, forming a periodic multistrip pattern, would one get oscillatory magnetic coupling or GMR effect?

As growing a laterally periodic multistrip pattern of, say Co/Cu, within a single layer would not be an easy task, we explore an alternative approach. We use a thin (a few nanometer) single layer of magnetic material, Ni, sandwiched between Si layers (Si/Ni/Si). In principle, a nanometer sized ion beam spot from a focused ion beam (FIB) source can be used to irradiate this system in a periodic strip pattern, expecting that in the irradiated strips there will be ion beam mixing of Si and Ni. The Ni-Si alloy (or compound) thus produced will serve as conducting nonmagnetic spacer strips for the periodic magnetic Ni strips. Should such patterned periodic multistrip system show an oscillatory magnetic coupling that would find applications like the magnetic multilayer systems. Albeit observation of such oscillatory magnetic coupling would require the width of the nonmagnetic strips of the order of a few nm, much smaller than what we attempted here. Even otherwise, such patterning would be a convenient way to fabricate lateral spin valve systems [3]. Here we investigate the ion irradiation induced modifications in a Si/Ni/Si system and utilize the knowledge for a magnetic/nonmagnetic multistrip patterning.

We begin with a Si/Ni/Si trilayer system. As currently available common FIB sources provide 30 keV Ga ions, we first use 30 keV Ga ions from a standard ion implanter for large area irradiation to study the evolution of magnetic behavior as a function of ion fluence and determine the threshold fluence where the magnetism is destroyed or ferromagnetic Curie temperature drops below room temperature. Once the threshold fluence is known, we use an FIB source to irradiate parallel strips on the trilayer at the threshold fluence. Thus we obtain a laterally periodic pattern of alternating magnetic (unirradiated) and paramagnetic/nonmagnetic (irradiated) strips.

2. Experimental

Si/Ni (*t*)/Si trilayer samples were grown in a molecular beam epitaxy (MBE) system. The thickness *t* of the Ni layer was chosen to be 10 nm, while the top Si layer was chosen to be 5 nm thick. Under ultrahigh vacuum condition clean Si(111) surfaces were produced. A Si buffer layer was first grown on the Si(111) substrate (100 mm dia). Ni was deposited on the buffer layer from an electron beam source with the substrate at room temperature. A thin (5 nm) amorphous Si layer was then deposited on the Ni layer of nominal thickness 10 nm and the samples were taken out of the UHV environment. Further measurements – magneto-optical Kerr rotation, Kerr microscopy and X-ray reflectivity – were carried out under ambient conditions. Rutherford backscattering spectrometry (RBS) experiments were also carried out on the as-prepared sample.

About 1 x 1 $cm^2$ samples were cut from the same wafer. Half of each sample was irradiated with 30 keV Ga ions from a low energy implanter with varying fluences from 1 x $10^{13}$ to 1 x $10^{15}$ ions/$cm^2$. The trilayer sample structure and the irradiation scheme are schematically shown in Fig.1.

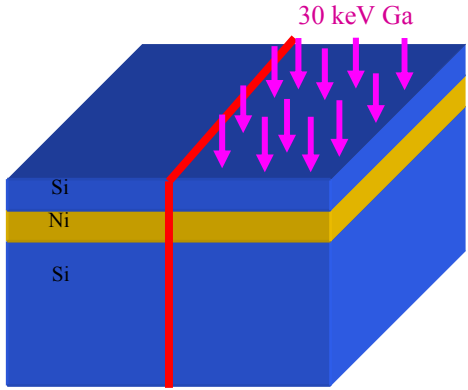

Fig. 1. The Si/Ni/Si trilayer and the ion irradiation scheme are shown schematically.

Measurements were made on both the irradiated and the unirradiated halves of each sample. The threshold fluence, at which the magnetic hysteresis loop at room temperature vanishes, has been identified.

Having identified the threshold fluence, 30 keV Ga ions from an Orsay FIB system (CANION 31Mplus, Orsay Physics) was used to irradiate parallel strips on the sample at this fluence in order to make a periodic magnetic-nonmagnetic multistrip pattern.

The integral magnetic characterization has been performed by longitudinal magneto-optic Kerr effect (MOKE) magnetometry. Helmholtz coils provide the applied magnetic field to avoid remanence effects and thus allow for an accurate determination of small magnetic fields. A diode laser with $\lambda = 405$ nm wavelength has been chosen in order to optimize the signal to noise ratio by maximizing the achievable Kerr rotation in the optical wavelength regime. The polarization rotation has been determined by two photodiodes sensing the two mutually orthogonal polarization directions in an optical bridge configuration.

The magnetic imaging of the microstructures was performed by magnetooptical Kerr microscopy in the longitudinal mode. The weak magneto-optical contrast was digitally enhanced by means of a background subtraction technique. The experimental setup has the option to apply in-plane magnetic fields in any direction independently of the magneto-optical sensitivity direction. To visualize the magnetic domains within the narrow stripes, the highest possible optical resolution, which is on the order of 300 nm for the given visible light illumination, was chosen.

X-ray reflectivity experiments were carried out at the ROBL beamline at ESRF, Grenoble. Using a Si(111) double crystal monochromator, 9.0 keV incident photons were selected and a point detector (scintillator) was used for the detection of the reflected X-rays. RBS experiments were carried out with 1 MeV $He^+$ ions from the 3 MV Pelletron accelerator at Institute of Physics, Bhubaneswar.

3. Results and Discussions

A. Rutherford backscattering spectrometry (RBS)

A RBS spectrum obtained by using 1 MeV $He^+$ ions from a pristine Si/Ni(10 nm)/Si sample is shown in Fig. 2. The trilayer structure is not evident from the spectrum due to insufficient resolution. However, the simulated spectrum for a Ni layer with sharp Ni/Si interfaces does not match with the measured spectrum. This indicates the presence of a Ni-Si alloy or reacted layer at the interfaces. This has important consequences for magnetism for the trilayer system with a thin Ni layer, as discussed in the next section.

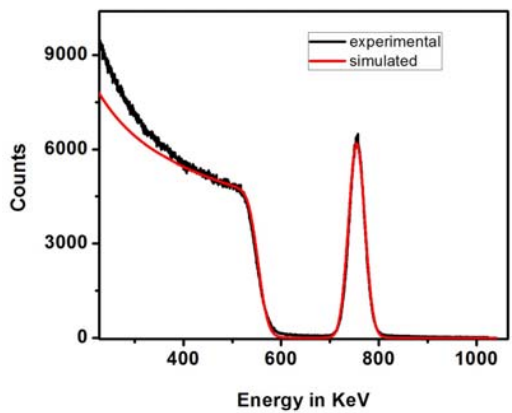

Fig. 2 A RBS spectrum from the Si/Ni/Si system along with a simulated spectrum.

The standard RBS technique, as used here, has a typical depth resolution of 20 nm. We have used a higher resolution ( ~ 0.1 nm) technique, namely X-ray reflectivity, to analyze the irradiated samples. These results are presented in Section C.

B. Magneto-optical Kerr rotation

The Si/Ni(10 nm)/Si sample shows a magnetic hysteresis loop in Kerr rotation measurement at room temperature which proves the ferromagnetic properties of the sample. For our intended investigations, this is the starting point. Several samples were cut from this wafer. Half of each sample, as illustrated in Fig. 1, was irradiated with a Ga ion beam at different fluences. MOKE measurements were made both on the pristine and the irradiated halves of each sample. In-plane angular dependent MOKE measurements for both virgin and irradiated samples show that the samples are isotropic in the film plane. The results of the MOKE measurements as a function of Ga irradiation fluence are shown in Fig. 3(a).We notice that the coercive field ($H_c$) decreases with increasing ion fluence ($\phi$) for the irradiated samples. When the incident laser beam spot is at the boundary between the pristine and the irradiated halves, signature of two hysteresis loops is observed, as shown in Fig. 3(b).The dependence of $H_c$ on $\phi$ is shown in Fig. 3(c). Fig. 3(c) also shows an inset containing a part of the present result along with the results of ion irradiation on a Pt/Co/Pt trilayer [2].The same trend of $\phi$-dependence of $H_c$ is observed in both cases. For the Pt/Co/Pt case, where 30 keV He ions were used, we notice that the fluence scale is about an order of magnitude higher compared to our case of 30 keV Ga ion irradiation of Si/Ni/Si. As we will see latter, ion beam induced atomic displacements is mainly responsible for the observed effect. 30 keV Ga ions are much more effective than 30 keV He ions in causing atomic displacements. That is why, a much less fluence of Ga ions is necessary to cause similar effects. For a fluence of 3 x $10^{14}$ ions/cm$^2$, there is still a small coercive field with a much reduced magnitude of Kerr rotation. This reduced Kerr magnitude is indicative of a reduced magnetic moment originating from a reduced magnetic active thickness due to ion irradiation induced interfacial mixing [5, 11, 12, 13]. At a fluence of 1 x $10^{15}$ ions/cm$^2$ the Kerr rotation signal is completely suppressed evidencing the loss of ferromagnetic properties at room temperature. This fluence is now taken as the threshold fluence for the destruction of ferromagnetism for the type and energy of ions used here. The multistrip pattern, fabricated by FIB irradiation, will be discussed in Section D.

We explore the reason for the ion beam induced gradual destruction of magnetism in the next section via X-ray reflectivity experiments. Here, let us try to understand the MOKE results in terms of what happens in the 30 keV Ga$^+$ interaction with the Si/Ni/Si trilayer system. Fig. 4(a) shows a SRIM simulation [14] displaying the distribution of displaced Ni and Si atoms. It is clear from the figure that in addition to Si being displaced in Si layers and Ni in the Ni layer, both Si and Ni get displaced across the interface into the neighboring layer(s). The concentration of the displaced atoms will increase with increasing ion fluence causing enhanced atomic mixing at the interfaces. So the effective thickness of the Ni layer would decrease. We also notice from Fig. 4(b) that a significant fraction of Ga ions are incorporated into the Ni layer. So, the dilution of Ni with incorporated Si and/or Ga may lead to a reduction in Curie temperature, which also leads to a loss of ferromagnetic properties, e.g. saturation magnetization [5], at room temperature. Of course, Ga concentration would be negligible, even if all of the $10^{15}$ ions are incorporated into the Ni layer. As we will see latter, at this fluence a 18.3 nm $Ni_{0.68}Si_{0.32}$ alloy layer is formed due to Si incorporation into the Ni layer. At the highest ion fluence, the Curie temperature might be below room temperature rendering the irradiated material paramagnetic at room temperature.

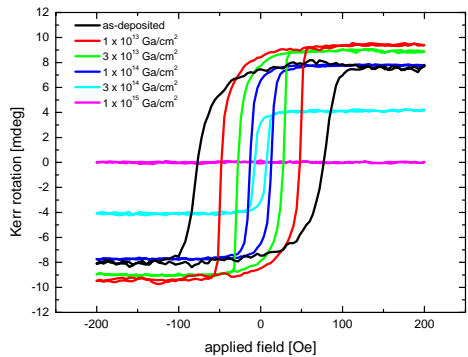

Fig. 3(a) Kerr rotation vs applied magnetic field, obtained from MOKE measurements on samples irradiated at different fluencies.

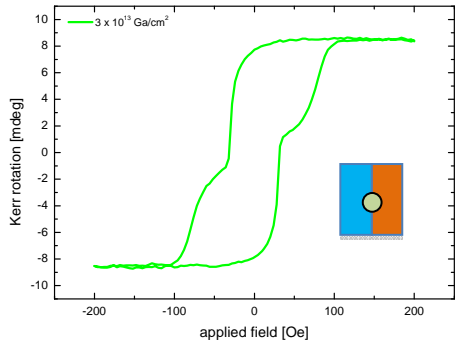

Fig. 3(b) MOKE measurement from an area at the boundary between the irradiated and the pristine part of a sample shows the signature of two hysteresis loops of different coercivities.

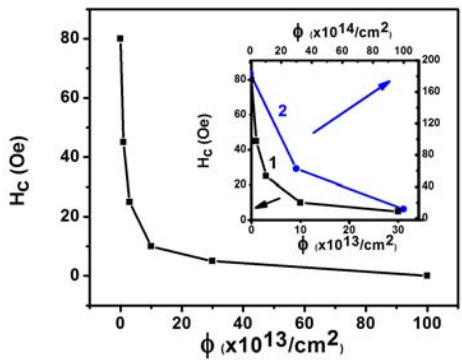

Fig. 3(c) Coercivity as a function of ion fluence. Curve 2 in the inset (from Ref. 2) shows a similar trend for ion irradiation in a Pt/Co/Pt trilayer.

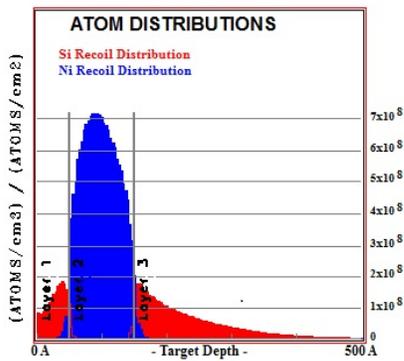

Fig. 4(a) Simulation showing atomic displacements across the interfaces in a Si(5 nm)/Ni(10 nm)/Si system. At the interfaces displaced Ni enter into Si and vice versa.

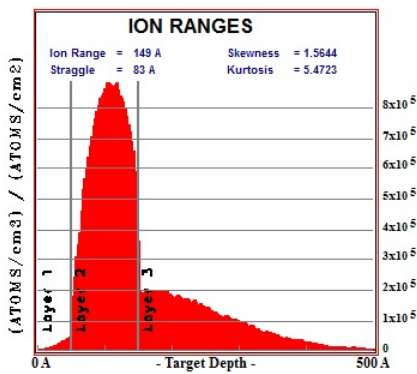

Fig. 4(b) Simulation showing the depth distribution of implanted Ga ions.

C. X-ray reflectivity

Although, in principle, the RBS technique can be used to study ion beam induced atomic displacements or intermixing in a layered structure, the typical depth resolution in a standard RBS measurement is about 20 nm. Thus for our system, Si(5nm)/Ni(10 nm)/Si, the standard RBS technique is not suitable for studying the microstructure. We use a high depth resolution (typically ~ 0.1 nm) technique, X-ray reflectometry, to investigate the microstructure. The evolution of X-ray reflectivity from the samples irradiated at increasing ion fluences is shown in Fig. 5.The reflectivity pattern from the pristine sample is typical from a sample containing two layers of different materials on a substrate (two frequency components). From the analysis of the reflectivity data, the thickness of the Ni layer is found to be 11.5 nm. For increasing fluences, the reflectivity patterns get gradually modified. Analysis shows results similar to that for the Si/Co/Si system [4], namely interface mixing and a gradual reduction of the Ni layer thickness. For the highest fluence, the reflectivity contains a single frequency oscillation, indicating the presence of a single layer of a homogeneous material on the silicon substrate. Detailed analysis shows that this is a layer of composition $Ni(1-x)Si(x)$ (with $x = 0.32$) and its thickness is 18.3 nm. The larger thickness compared to the original thickness of the Ni film is due to alloying with Si. This alloying explains why loss of ferromagnetism is observed in this irradiated sample. In the Ni-Si system, ferromagnetic Curie temperature ($T_c$) strongly reduces with small increases of the Si content [15]. For pure Ni, $T_c$ is 354.4 °C. Even for 10% Si incorporation, the value of $T_c$ goes down below 0 °C. According to Ref. 14, $T_c$ is -200 °C for a Si content of ~13%. For the layer composition, produced in ion irradiation at the threshold fluence, the $T_c$ value is apparently far below room temperature. Once the threshold fluence is determined and the reason for the loss of ferromagnetism is understood, we proceed with the objective of fabricating a periodic magnetic-nonmagnetic laterally patterned multistrip structure.

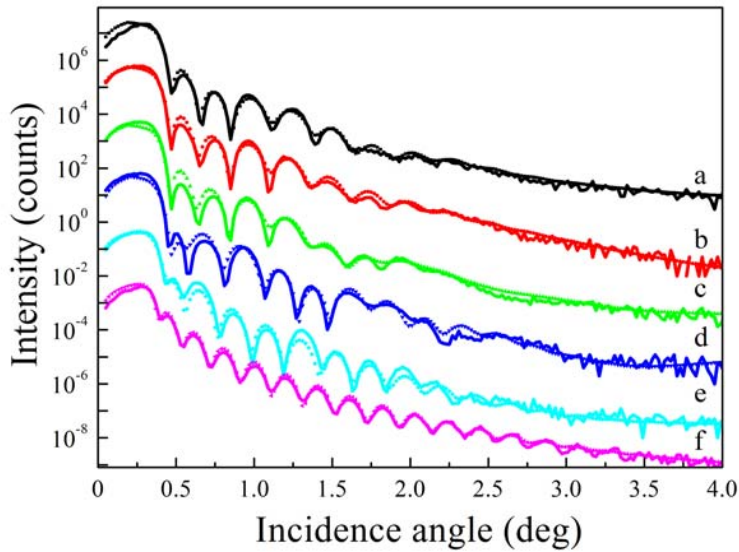

Fig. 5 X-ray reflectivity results from irradiated samples for different ion fluences and the corresponding simulated curves: (a) pristine, (b) $1 \times 10^{13}$, (c) $3 \times 10^{13}$, (d) $1 \times 10^{14}$, (e) $3 \times 10^{14}$, (f) $1 \times 10^{15}$. Successive curves are vertically shifted by two orders of magnitude for clarity.

D. Patterning with a focused ion beam

A laterally periodic multistrip pattern fabricated by using a 30 keVGa ion beam from an FIB source is shown in Fig.6. The smallest beam spot (FWHM) of our FIB system is 20 nm. In order to observe any oscillatory magnetic coupling in such multistrip structure, one perhaps needs a much smaller width of the nonmagnetic strips and hence a much smaller beam spot from an FIB source. However, this study represents a preliminary effort to fabricate a multistrip pattern. The multistrip pattern shown here, Fig. 6(b), has 50 nm wide irradiated strips while the width of the unirradiated strips is 1 µm. The image in Fig. 6(b) was generated by collecting emitted secondary electrons.

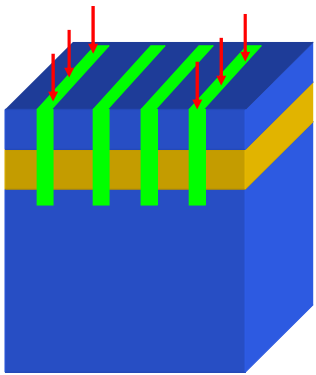

Fig. 6(a) Patterning with FIB is schematically shown. Ion irradiation was carried out along the stripes. Ion beam induced modification extends from the top Si layer, through the Ni layer, into a part of the substrate Si.

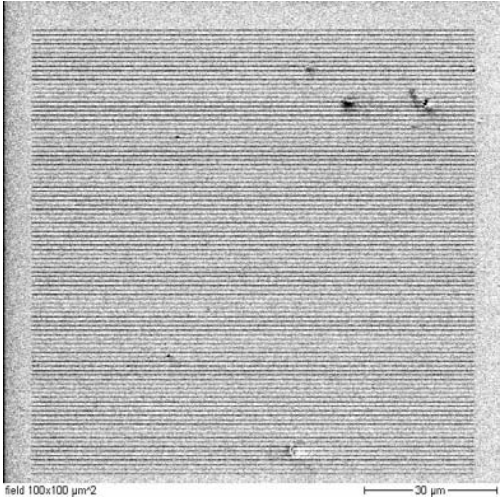

Fig. 6(b) A multistrip pattern generated by FIB patterning.

E. Kerr microscopy

The magnetic domain structures after the application of magnetic fields either parallel or perpendicular to the stripe axis are displayed in Fig. 7. Whereas the magnetization in the surrounding Ni film structure displays signatures of a magnetic isotropic film, the high contrast domains obtained for both field directions indicate a magnetization alignment perpendicular to the stripe axis. In contrast to simple arguments, considering dominating shape anisotropy effects which would favor an alignment of magnetization parallel to the stripe axis, the opposite behavior is observed. At this point the exact reason for the occurrence for the interacting domains across several stripes is not known, but magneto-elastic effects introducing a stress-induced anisotropy component oriented perpendicular to the stripe axis are the most plausible explanation for the observed behavior.

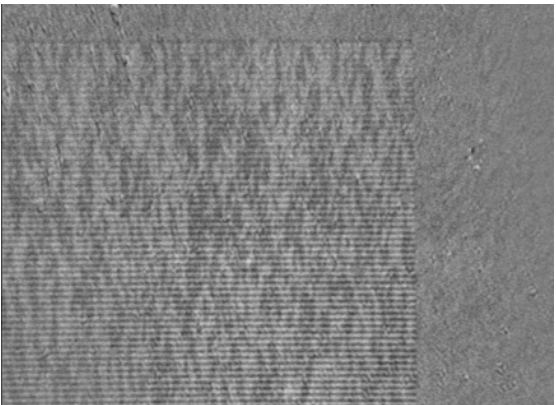

Fig. 7(a)

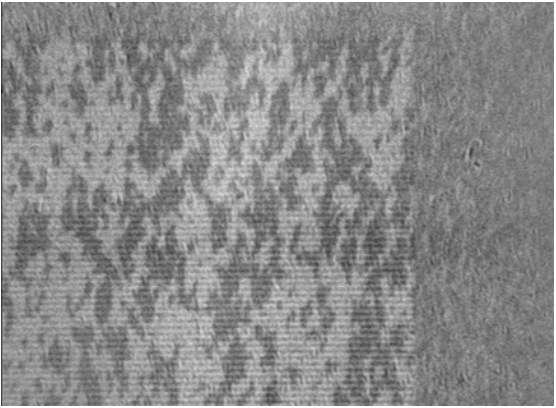
Fig. 7(b)

Fig. 7. Kerr microscopy images displaying the magnetic domain structure after applying a magnetic field (a) parallel and (b) perpendicular to the stripe axis.

4. Summary and Conclusions

With the idea of reducing the dimension of a magnetic multilayer system, we attempted to fabricate a magnetic multistrip system from a single ferromagnetic layer by destroying ferromagnitsm at room temperature in alternating stripes on this layer. For this purpose we have used a Ni layer sandwiched between Si layers (Si/Ni/Si) and used 30 keV Ga ion irradiation from a focussed ion beam (FIB) source. This has produced irradiated stripes of 50 nm width, seperated by 1 μm unirradiated stripes. The ion fluence for FIB irradiation has been chosen from control experiments using a 30 keV Ga ion beam from a standard ion source. From these experiments, a threshold fluence has been identified so that ferromagnetism at room temperature is lost at this ion fluence. Following FIB ion irradiation at this threshold ion fluence, the unirradiated stripes remain ferromagnetic at room temperature while the irradiated stripes, based on the control experiments, would be either paramagnetic at room temperature or nonmagnetic. However, MOKE microscopy experiments show that the intended magnetic/nonmagnetic multistrip pattern has not formed. Considering shape anisotropy effects, which would favor an alignment of magnetization parallel to the stripe axis, the opposite behavior is observed. Interacting magnetic domains across several stripes are observed. Magneto-elastic effects introducing a stress-induced anisotropy component oriented perpendicular to the stripe axis are the most plausible explanation for the observed behavior.

Acknowledgement

We thank F. Allenstein and G. Beddies for providing the as-grown sample. We also thank accelerator staff at Institute of Physics for help with the RBS measurement.

*msbnd@iacs.res.in
#Retired and pensioner
$ Present address: Helmholtz-Zentrum Geesthacht, Institute of Materials Research, Max-Planck-Straße 1, 21502 Geesthacht, Germany